\documentclass[twocolumn,showpacs,aps,pra,floatfix,superscriptaddress]{revtex4-1}
\usepackage{amssymb,amsthm, amssymb, latexsym}
\usepackage{bbm}
\usepackage{graphicx}
\usepackage{amsmath}
\usepackage[english]{babel}
\usepackage{pstricks}

\begin{document}
\title{U(3) artificial gauge fields for cold atoms}

\author{Yu-Xin Hu}
\affiliation{Centre for Quantum Technologies, National University of Singapore, 3 Science Drive 2, Singapore 117543, Singapore}
\affiliation{Merlion MajuLab, CNRS-UNS-NUS-NTU International Joint Research Unit UMI 3654, Singapore}

\author{Christian Miniatura}
\affiliation{Merlion MajuLab, CNRS-UNS-NUS-NTU International Joint Research Unit UMI 3654, Singapore}
\affiliation{Institut Non Lin\'{e}aire de Nice, UMR 7335, UNS, CNRS; 1361 route des Lucioles, 06560 Valbonne, France}
\affiliation{Centre for Quantum Technologies, National University of Singapore, 3 Science Drive 2, Singapore 117543, Singapore}
\affiliation{Department of Physics, National University of Singapore, 2 Science Drive 3, Singapore 117542, Singapore}
\affiliation{Institute of Advanced Studies, Nanyang Technological University, 60 Nanyang View, Singapore 639673, Singapore}

\author{David Wilkowski}
\affiliation{Merlion MajuLab, CNRS-UNS-NUS-NTU International Joint Research Unit UMI 3654, Singapore}
\affiliation{Institut Non Lin\'{e}aire de Nice, UMR 7335, UNS, CNRS; 1361 route des Lucioles, 06560 Valbonne, France}
\affiliation{Centre for Quantum Technologies, National University of Singapore, 3 Science Drive 2, Singapore 117543, Singapore}
\affiliation{School of Physical and Mathematical Sciences, Nanyang Technological University, Singapore 637371, Singapore}

\author{Beno\^{\i}t~Gr\'{e}maud}
\affiliation{Merlion MajuLab, CNRS-UNS-NUS-NTU International Joint Research Unit UMI 3654, Singapore}
\affiliation{Laboratoire Kastler Brossel, Ecole Normale Sup\'{e}rieure CNRS, UPMC; 4 Place Jussieu, 75005 Paris, France}
\affiliation{Centre for Quantum Technologies, National University of Singapore, 3 Science Drive 2, Singapore 117543, Singapore}
\affiliation{Department of Physics, National University of Singapore, 2 Science Drive 3, Singapore 117542, Singapore}

\date{\today}

\begin{abstract}
We propose to generate an artificial non-Abelian $U(3)$ gauge field by using a 2-tripod scheme, namely two tripod configurations sharing a
common ground state level and driven by resonant 1-photon transitions. Using an appropriate combination of four Laguerre-Gauss and
two Hermite-Gauss laser beams, we are able to produce a $U(3)$-monopole and a $U(3)$ spin-orbit coupling for both alkali and alkaline-earth atoms.
This 2-tripod scheme could open the way to the study of interacting spinor condensates subjected to $U(3)$-monopoles.
\end{abstract}


\maketitle

\section{Introduction}

Within less than a decade, ultracold quantum gases have successfully pervaded many fields of physics.
Indeed, they provide a rather unique testing bed where theorists' dreams can be turned into carefully designed experimental situations.
This is particularly true in the condensed matter realm where they became a key player in many-body physics~\cite{Lewenstein07,Blochreview08,Ketterle2}.
Quantum Hall effects did not escape the trend. The catch however is that atoms are neutral and one thus needs to implement an artificial gauge
field acting on the atoms that would give rise to a strong enough effective magnetic field. A first idea was to set quantum gases into rapid
rotation~\cite{Cooper2008}. Since then, more versatile and promising schemes have been introduced, some even realized, all based on
light-atom interactions~\cite{Spielman09a,Spielman09b,Spielman11,Zhang12,Zwierlein12,Windpassinger12,Spielman12}. These light-induced artificial
gauge fields, encompassing Abelian and non-Abelian situations, have opened the door to a whole class of model
Hamiltonians~\cite{Dalibard11,Cooper2011,Juzeliunas2012,Goldman2013} and are addressing diverse physical situations ranging
from artificial Dirac monopoles~\cite{Ray2014}, spin-orbit (SO) coupling~\cite{Zhu2006,Stanescu2008,Barnett2012} and
topological phases~\cite{Mottonen09,Bercioux11}, to non-Abelian particles~\cite{Burrello10},
and mixed dimensional systems~\cite{Nishida_2008,Nishida_2010,Lamporesi_2010,Huang_2013,Iskin_2010}.

In this paper, we discuss a new proposal to generate artificial non-abelian $U(3)$ gauge fields.
Our scheme, based on a single particle approach, is a straightforward generalization of the tripod scheme discussed in~\cite{Ruseckas05}.
It is based on three space-dependent dark states arising from the coupling with resonant one-photon transitions between Zeeman sub-levels
belonging to different hyperfine states of
an alkali atom, such as $^{87}$Rb, subjected to a magnetic field. In the following, we first introduce
the laser scheme we propose and work out the general expressions for both the effective vector and scalar fields.
We next discuss two specific laser configurations: the first one gives rise to a non-Abelian $U(3)$ monopole while the second one
gives rise to a non-Abelian SO-like coupling. Finally, we discuss alkaline-earth atoms, taking the fermionic isotope of Strontium as
a paradigmatic example. In this case however, because the Zeeman shifts of the lowest hyperfine states $^1\!{S}_0$ are negligible, a
slightly different laser configuration is required to appropriately couple the electronic levels.

In both situations, monopole or spin-orbit coupling, working with a gauge group larger than $U(2)$, 
brings potentially more interesting
physics for two main reasons. First, the number of species being larger, new and nontrivial many-body phenomena can emerge. 
For example, it has been found that a 3-color fermionic system with attractive interactions gives rise to a nontrivial trionic 
ground state separating the usual BEC and BCS phases~\cite{Floerchinger09}. Second, the gauge group being larger, it contains 
more subgroups and these subgroups can still have a nontrivial structure. Therefore, in the presence of interactions, a larger 
gauge group allows for different symmetry-breaking scenarios for the ground state. 
For instance, in the situation of a Higgs field coupled to a $U(3)$ gauge field, i.e. extending the 't~Hooft monopole 
to a larger group, it has been found that
the $U(3)$-monopole solutions can have two different kinds of topology, 
depending on the subgroup leaving the ground state invariant~\cite{Sinha76}.
Similarly,
it has been shown, that, on a square lattice, a system with a $U(3)$ spin-orbit coupling, even in the non-interacting regime,
has topologically nontrivial states, in contrast to the $U(2)$ case~\cite{Barnett2012}.

\section{2-tripod scheme}

\begin{figure}[tb]
\centering
\includegraphics*[width=8cm]{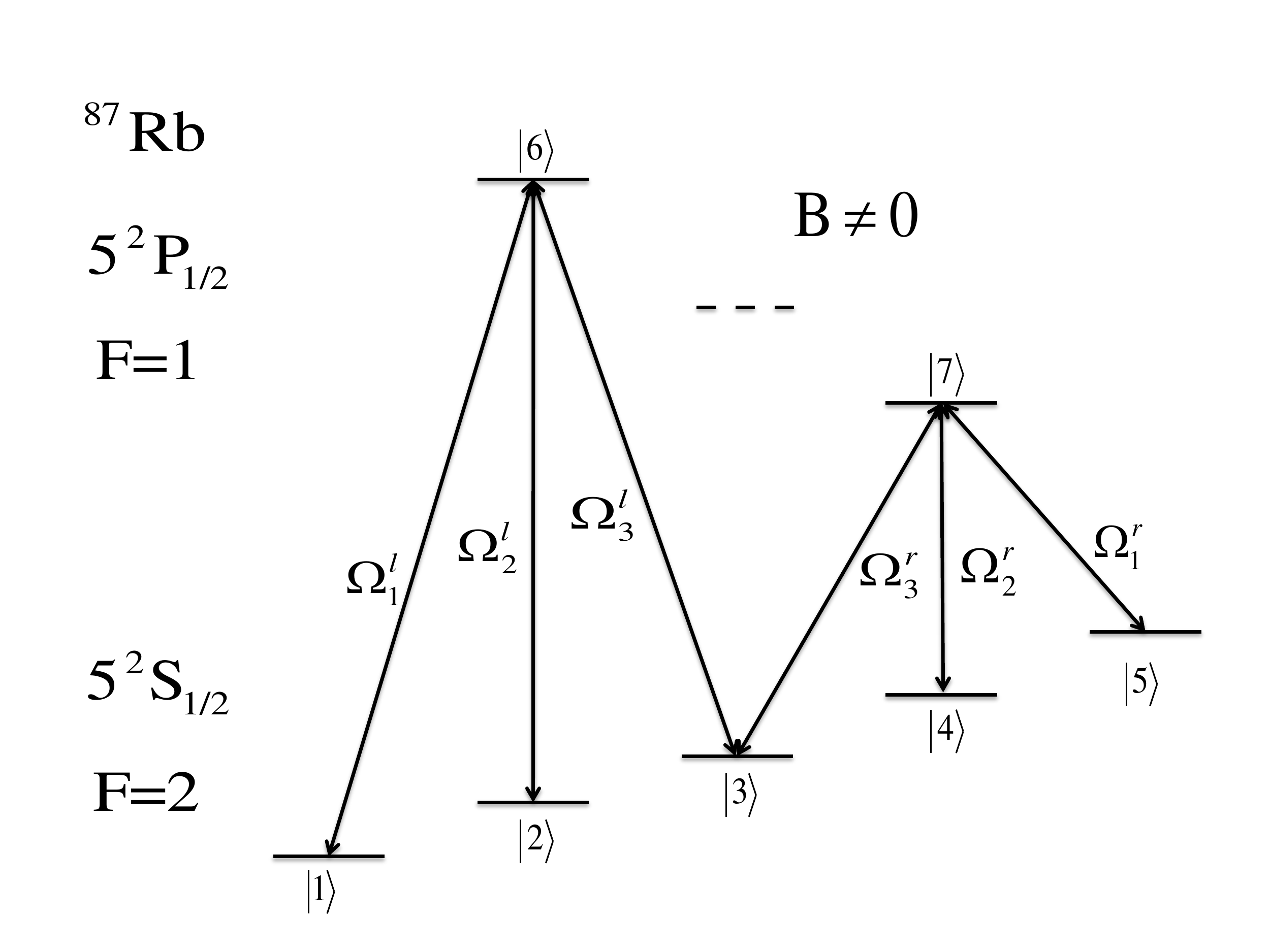}
\caption{\label{dtripod} 2-tripod scheme for the $D_1$ line of $^{87}$Rb.
An external magnetic field, chosen as the quantization axis, is applied to the atoms and lifts the Zeeman degeneracy of the ground state
(with total spin $F_g=2$) and excited state (with total spin $F_e=1$) manifolds.
Six resonant laser beams then illuminate the atoms,
their polarization state being appropriately chosen to address the $\pi$ and $\sigma_\pm$ transitions shown in the Figure.
States $\vert i \rangle$ ($1 \leq i \leq 5$) refer to the Zeeman levels $\vert 2, m_g\rangle$ ($|m_g| \leq 2$) in the ground state manifold.
States $\vert i \rangle$ ($i=6,7$) refer to the Zeeman levels $\vert 1, m_e=\pm 1 \rangle$ of the excited states manifold.
All other excited levels (three of them indicated by horizontal dashed lines) are not addressed since they are
far off-resonance from any of the six laser beams. This 2-tripod scheme gives rise to three dark states with vanishing energy.
}
\end{figure}

We consider the 2-tripod coupling scheme depicted in Fig.~\ref{dtripod}. It is based on two usual tripod schemes. One couples ground states $\vert 1\rangle$, $\vert 2 \rangle$, $\vert 3 \rangle$ to excited level $\vert 6\rangle$. The other one couples ground states $\vert 3\rangle$, $\vert 4 \rangle$, $\vert 5 \rangle$ to the excited level $\vert 7\rangle$. As one can see these two tripod schemes are
not independent since they share one common ground state level, namely level $\vert 3\rangle$.
This very situation can be implemented with alkali atoms, for instance by considering the $D_1$ line of $^{87}$Rb atoms.
In this case, one first applies a magnetic field to split the Zeeman structure of both the ground $F_g=2$ and excited $F_e=1$ states and
then one shines six suitably polarized resonant laser beams to produce the desired 2-tripod coupling scheme shown in Fig.~\ref{dtripod}.
Sec.~\ref{section_Possible_exp_realization} below gives more  details about the experimental realization and its limitations.

Since the 2-tripod scheme involves five ground states coupled to two excited states, one expects three degenerate dark
states (DS) with vanishing energy. Indeed, in the rotating wave approximation, the 2-tripod Hamiltonian reads:
\begin{align}
H_{0}=&-\hbar\big(\Omega_{1}^{l}\vert 6 \rangle \langle 1 \vert+\Omega_{2}^{l}\vert 6 \rangle \langle 2 \vert+
             \Omega_{3}^{l}\vert 6 \rangle \langle 3 \vert\nonumber \\
             &\quad+\Omega_{3}^{r}\vert 7 \rangle \langle 3 \vert+
             \Omega_{2}^{r}\vert 7 \rangle \langle 4 \vert+\Omega_{1}^{r}\vert 7 \rangle \langle 5 \vert\big)+h.c.
\end{align}
We now parametrize the position-dependent Rabi frequencies as follows~\cite{Ruseckas05,Dalibard11}:
\begin{equation}
 \begin{aligned}
 & \Omega_{1}^{a}=\Omega_{a} \sin\theta_{a} \cos\phi_{a} e^{iS^{a}_{1}}\\
 & \Omega_{2}^{a}=\Omega_{a} \sin\theta_{a} \sin\phi_{a} e^{iS^{a}_{2}}\\
 & \Omega_{3}^{a}=\Omega_{a} \cos\theta_{a} e^{iS^{a}_{3}},
 \end{aligned}
\end{equation}
where $a=l,r$. The twelve different quantities $\Omega_{i}^{a}$, $\theta_{a}$, $\phi_{a}$ and $S_{i}^{a}$ ($i=1,2,3$) are generally space-dependent.
It is then straightforward  to compute the three orthonormal DS of the 2-tripod scheme:

\begin{subequations}
\label{DS3}
\begin{align}
  |D_{l}\rangle =& \sin\phi_{l} e^{iS^{l}_{31}}|1\rangle - \cos\phi_{l} e^{iS^{l}_{32}}|2 \rangle\\
 |D_{r}\rangle =& \sin\phi_{r} e^{iS^{r}_{31}}|5\rangle - \cos\phi_{r} e^{iS^{r}_{32}}|4\rangle\\
\vert D_{0} \rangle =&\frac{1}{\alpha_0}\left[ \cot\theta_{l} \cos\phi_{l} e^{iS^{l}_{31}}\vert 1 \rangle +
                      \cot\theta_{l}\sin\phi_{l}e^{iS^{l}_{32}}\vert 2\rangle - \vert 3 \rangle\right. \nonumber \\
                      &\left.+\cot\theta_{r} \cos\phi_{r} e^{iS^{r}_{31}}\vert 5 \rangle
                       +\cot\theta_{r}\sin\phi_{r}e^{iS^{r}_{32}}
                       \vert 4\rangle \right] \nonumber\\
 \alpha_0 =& (1+\cot^{2}\theta_{l}+\cot^{2}\theta_{r})^{1/2}
 \end{align}
 \end{subequations}
where $S_{ij}^{a}=S_{i}^{a}-S_{j}^{a}$ ($a=l,r$).
One may note that when $\Omega_{3}^{r}=0$, corresponding to $\theta_r=\pi/2$, then the 2-tripod scheme breaks up into a $U(2)$
tripod configuration coupling states $\vert 1\rangle$, $\vert 2 \rangle$, $\vert 3 \rangle$ to $\vert 6\rangle$ and an
independent $U(1)$ $\Lambda$-configuration coupling states $\vert 4 \rangle$, $\vert 5 \rangle$ to $\vert 7\rangle$.
Then state $\vert D_r\rangle$ identifies with the DS of the $\Lambda$-configuration while $\vert D_l\rangle$ and the
corresponding $\vert D_0\rangle$ state identify with the two DS of the left-tripod configuration. The same type of
considerations can be made if $\Omega_{3}^{l}=0$, corresponding to $\theta_l=\pi/2$. In other words, the
states $\vert D_a\rangle$ ($a=l,r$) are DS for the left and right $U(2)$ tripod configuration as well as DS for the $\Lambda$ configuration.
The remaining state $|D_0\rangle$, embodying all five Zeeman ground state levels, reflects the coupling of the two $U(2)$ tripod
configurations when both $\Omega_{3}^{l}$ and $\Omega_{3}^{r}$ are non zero. It boils down to the missing tripod DS
when $\Omega_{3}^{l}$ or $\Omega_{3}^{r}$ vanishes.

\begin{figure}[tb]
\centering
\includegraphics*[width=8cm]{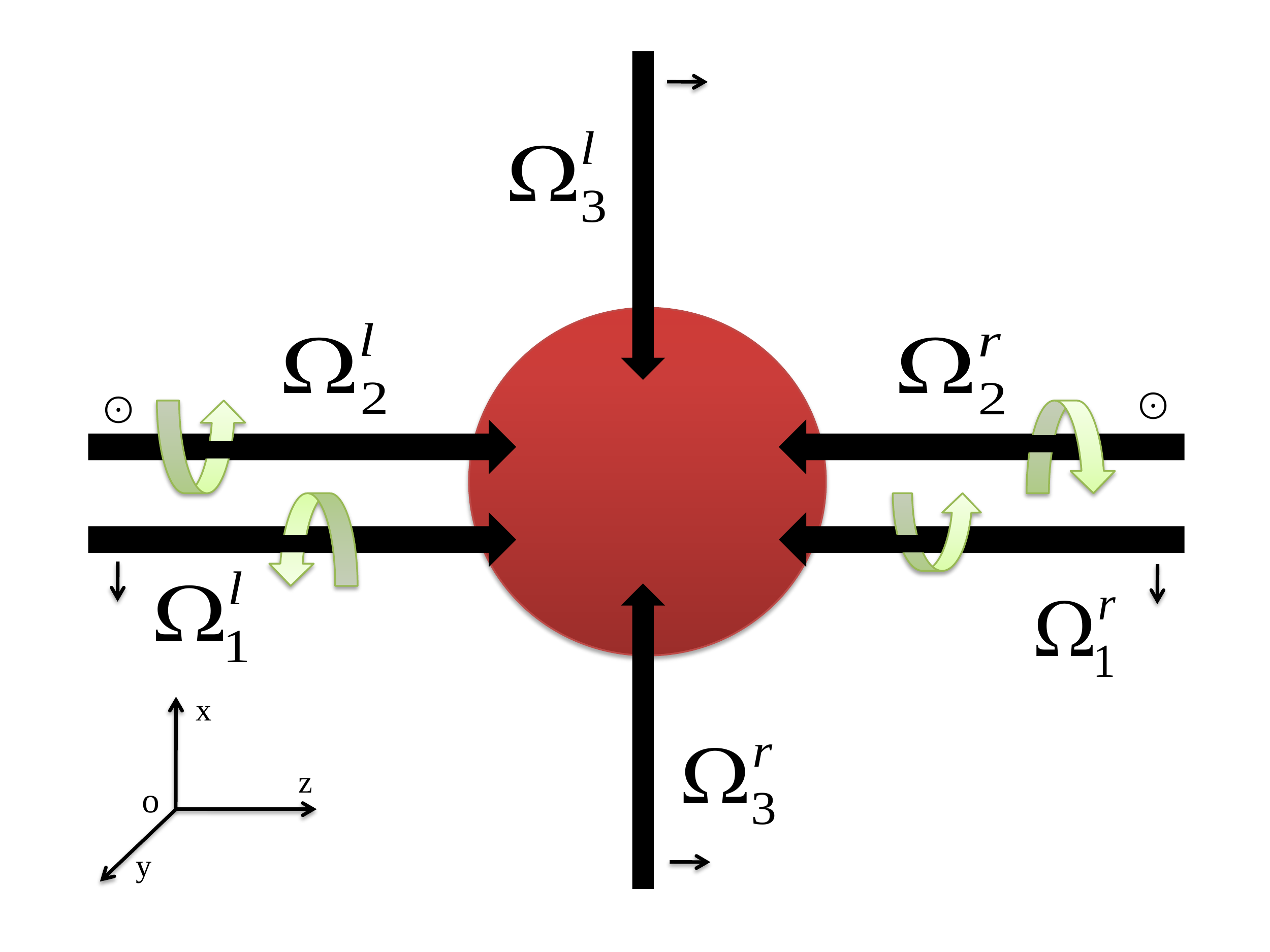}
\caption{\label{su3sl} [Color online] Laser beam configuration giving rise to a non-Abelian $U(3)$-monopole with unit charge and associated with the generator $J_x$ of the $SO(3)$ subgroup. The horizontal beams are
Laguerre-Gauss modes carrying opposite orbital angular momentum $\pm\hbar$ shown by the green arrows.
The vertical laser beams are $n=1$ Hermite-Gauss modes.
The bias magnetic field lifting the Zeeman degeneracies in Fig.~\ref{dtripod} defines the quantization axis.
One possible configuration consists in choosing the magnetic field along $Oy$. Then, all beams are linearly-polarized (thin black arrows) and can selectively address their allocated transitions since their polarization state have a non-vanishing projection on the desired $\sigma_\pm$ and $\pi$ transitions.
}
\end{figure}

From the DS expressions~\eqref{DS3}, one can derive the vector and scalar potentials associated to the 2-tripod scheme. The vector potential $\vec{\mathbf{A}}$ is now a $3\times3$ Hermitian matrix with entries:
\begin{equation}
\label{vecA}
\begin{aligned}
\vec{A}_{11}=&\cos^{2}\phi_{l}\vec{\nabla} S^{l}_{23} + \sin^{2}\phi_{l}\vec{\nabla} S^{l}_{13}\\
\vec{A}_{33}=&\cos^{2}\phi_{r}\vec{\nabla} S^{r}_{23} + \sin^{2}\phi_{r}\vec{\nabla} S^{r}_{13}\\
\vec{A}_{22}=&\frac{1}{\alpha_0^2}\biggl[ \cot^2\theta_{l}(\cos^{2}\phi_{l}\vec{\nabla} S^{l}_{13}+\sin^{2}\phi_{l}\vec{\nabla} S^{l}_{23})\biggr.\\
&\biggl.+\cot^2\theta_{r}(\cos^{2}\phi_{r}\vec{\nabla} S^{r}_{13}+\sin^{2}\phi_{r}\vec{\nabla} S^{r}_{23})\biggr]\\
\vec{A}_{21}=&\vec{A}^{*}_{12} = \frac{\cot\theta_{l}}{\alpha_0}\left(\frac{1}{2}\sin(2\phi_{l})\vec{\nabla} S^{l}_{12}+i \vec{\nabla}\phi_{l}\right)\\
\vec{A}_{23}=&\vec{A}^{*}_{32} = \frac{\cot\theta_{r}}{\alpha_0}\left(\frac{1}{2}\sin(2\phi_{r})\vec{\nabla} S^{r}_{12}+i \vec{\nabla} \phi_{r}\right)\\
\vec{A}_{31}=& \vec{A}^{*}_{13} = \,0,
\end{aligned}
\end{equation}
where the star denotes complex conjugation. The scalar potential expression is rather involved and is given in the Appendix~\ref{scalpot} for sake of completeness.

\section{$U(3)$ monopole}
\label{Section_U3_monople}

In the following, we will use the spherical coordinate system $(r, \theta,\varphi)$ about axis $Oz$ to parametrize a point $M(x,y,z)$ in space. Then, from Eq.~\eqref{vecA},
one can check that a gauge field corresponding to a $U(3)$ monopole can be generated by using the following Rabi frequencies:
\begin{equation}
\label{u3jx}
 \begin{aligned}
\Omega_{1,2}^{l}=&\Omega_{0}\frac{\rho}{R} e^{i(kz\mp \varphi)} \qquad \Omega_{3}^{l}=\Omega_{0}\frac{z}{R}e^{ikx}\\
\Omega_{1,2}^{r}=&\Omega_{0}\frac{\rho}{R} e^{-i(kz\pm \varphi)} \qquad \Omega_{3}^{r}=\Omega_{0}\frac{z}{R}e^{-ikx},
\end{aligned}
\end{equation}
where $\rho = \sqrt{x^2+y^2} = r \sin\theta$. The corresponding laser beam configuration is shown in Fig.~\ref{su3sl},
the quantization axis being along axis $Oy$. The three beams addressing the left tripod configuration consist of two linearly-polarized
co-propagating (along axis $Oz$) Laguerre-Gauss beams with orbital angular momentum $\pm \hbar$ and of a linearly-polarized Hermite-Gauss beam
propagating along axis $Ox$.
The three remaining beams are just "reflection images" of the previous beams and address the right tripod configuration.
The potential vector then reads:
\begin{equation}
\begin{aligned}
\vec{\mathbf{A}}=&-\hbar \frac{\cos\theta}{\sqrt{2}\sin\theta}\frac{\hat{e}_{\varphi}}{r}\left(
                      \begin{array}{ccc}
                        0 & 1 & 0 \\
                        1 & 0 & 1\\
                        0 & 1& 0 \\
                      \end{array}
                   \right)\\
 &-\hbar k\sin^{2}\theta \ (\hat{e}_{z}-\hat{e}_{x})\left(
                      \begin{array}{ccc}
                        0 & 0 & 0 \\
                        0 & 1 & 0\\
                        0 & 0 & 0 \\
                     \end{array}
                    \right)\\
&+\hbar k \, (\hat{e}_{z}-\hat{e}_{x}) \, \mathbbm{1}
\end{aligned}
\label{eq_A_Monople}
\end{equation}
The last term is inessential: it is a constant gradient term proportional to the unit matrix that can be gauged away through a
$U(1)$ transformation. The second term depends on the wave number $k$ and is similar to the $k$-dependent terms found in the $U(2)$
monopole case~\cite{Ruseckas05,Dalibard11}. It is not singular and leads to non-monopole terms. We will not discuss it in the following though
it can play an important role in the dynamics~\cite{Mottonen09}. Finally, the first term can be rewritten as $\vec{\mathbf{A}}_m = \vec{a}_m \, J_x$
where $\vec{a}_m = - \cos\theta \, \hat{e}_{\varphi} /(r\sin\theta)$. It corresponds to a non-Abelian $U(3)$ monopole with unit effective magnetic
charge $Q=1$ coupled to the spin-1 operator
\begin{equation}
J_x= \frac{\hbar}{\sqrt{2}} \,\left(
\begin{array}{ccc}
0 & 1 & 0 \\
 1 & 0 & 1\\
0 & 1& 0 \\ \end{array} \right).
\end{equation}
Indeed, the corresponding non-Abelian magnetic field $\vec{\mathbf{B}}_m = \vec{\nabla} \times \vec{\mathbf{A}}_m + \vec{\mathbf{A}}_m \times \vec{\mathbf{A}}_m /(\mathrm{i}\hbar)$ reads $\vec{\mathbf{B}}_m = (\vec{\nabla} \times \vec{a}_m) \, J_x$.
Let us first consider a general Abelian vector field of the form $\vec{a} = g(\theta) \, \hat{e}_{\varphi} /(r\sin\theta)$.
Then the corresponding Abelian magnetic field reads:
\begin{align}
&\vec{B} = \vec{\nabla} \times \vec{a} = \frac{g'(\theta)}{\sin{\theta}}\frac{\hat{e}_r}{r^2} - 2\pi \phi(z) \delta(x)\delta(y) \, \hat{e}_z \label{eq:AbMono} \\
& \phi(z) = g(0) \Theta(z) + g(\pi) \Theta(-z), \label{eq:StringFlux}
\end{align}
where $\Theta(u)$ is the step function. It consists of a monopole contribution $B_m$ given by the first term in the right-hand side
of Eq.\eqref{eq:AbMono} and of a Dirac-like string contribution with flux $-2\pi\phi(z)$. The magnetic charge $Q_m$ associated with the
monopole field is computed with the help of Gauss theorem. It reads:
\begin{equation}
 Q_m= \frac{1}{4\pi} \, \int_{S}\vec{B}\cdot \vec{dS} =  \frac{g(\pi)-g(0)}{2},
 \label{eq_charge}
\end{equation}
where $S$ is the sphere of radius $r$ and $\vec{dS}=r^2\sin\theta \, d\theta d\varphi  \, \hat{e}_r$~\cite{stringcharge}.
Since we have $g(\theta)=-\cos{\theta}$ in our 2-tripod situation, we thus get a genuine monopole field with unit magnetic charge coupled to $J_x$.
It is worth noting that, in high-energy physics, it has been established that \emph{all} non-Abelian monopoles are simply obtained as Abelian monopoles
times a \emph{constant} (charge) matrix $Q$~\cite{Coleman81}. The string is undetectable when $\exp(4\pi \mathrm{i} Q/\hbar) = \mathbbm{1}$.
This is also what we get here.

\begin{figure}[tb]
\centering
\includegraphics*[width=8cm]{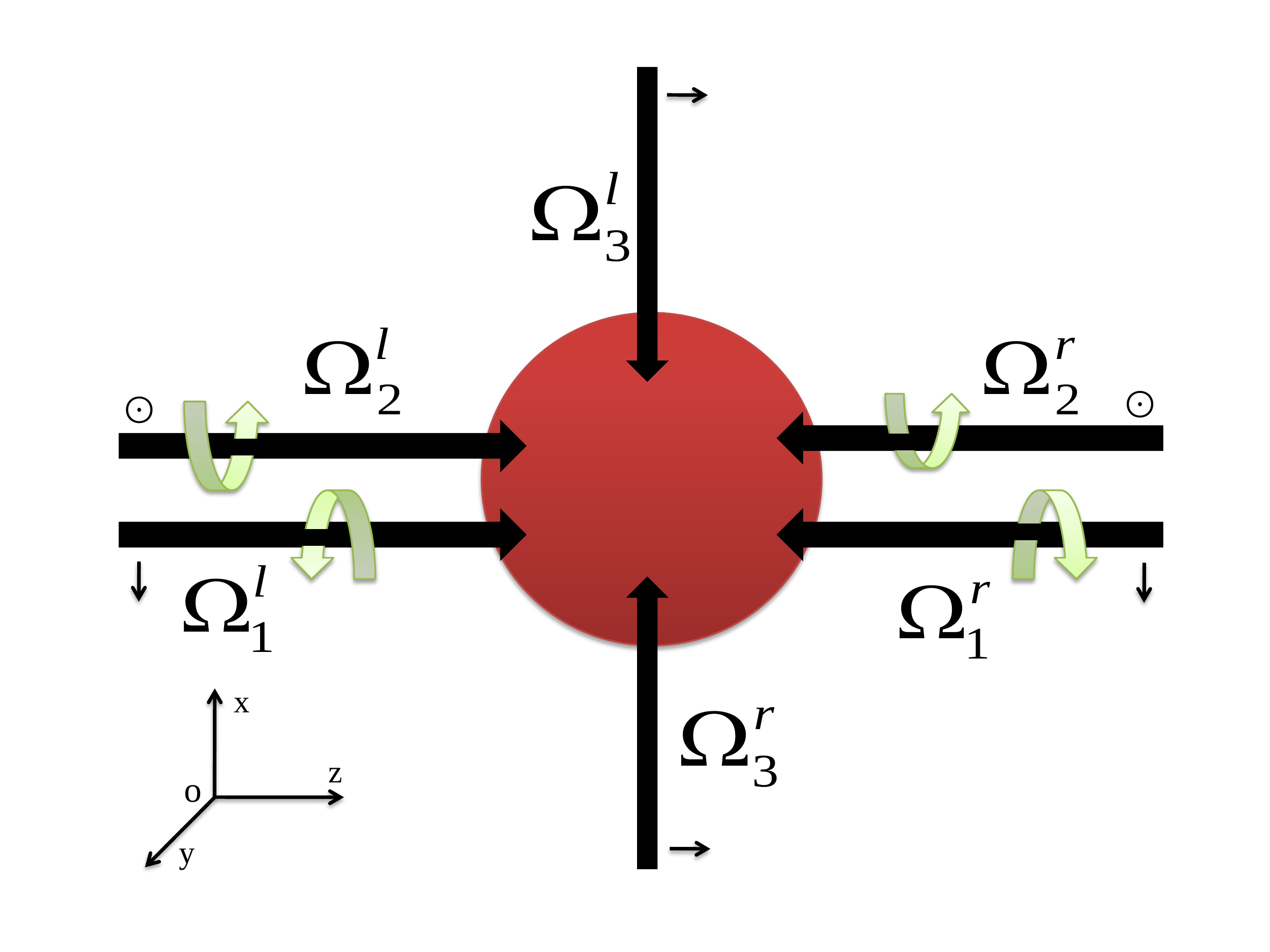}
\caption{\label{su3nsl} [Color online] Laser beams configuration generating a non-Abelian $U(3)$-monopole with unit
charge and associated with a $3\times 3$ matrix which does not belong to the $SO(3)$ subgroup. It is obtained by flipping
the sign of the orbital angular momentum carried by the right horizontal beams. The conventions and polarizations of the
beams are the same as in Fig.~\ref{su3sl}.
}
\end{figure}

If, starting from the laser configuration shown in Fig.~\ref{su3sl}, one just flips the sign of the orbital angular momentum carried by each of the right Laguerre-Gauss fields, i. e.
\begin{equation}
 \begin{aligned}
 \label{u3njx}
\Omega_{1,2}^{r}=&\Omega_{0}\frac{\rho}{R} e^{-i(kz\mp \varphi)},
\end{aligned}
\end{equation}
while keeping all the other fields unaffected (see Fig~\ref{su3nsl}), then only the monopole part of the full $U(3)$ vector potential
is modified and now reads $\vec{\mathbf{A}}_m= \, \vec{a}_m \  \tilde{J}_x$ where:
\begin{equation}
\begin{aligned}
                     \tilde{J}_x = \frac{\hbar}{\sqrt{2}} \, \left(
                      \begin{array}{ccc}
                        0 & 1 & 0 \\
                        1 & 0 & -1\\
                        0 & -1& 0 \\
                      \end{array}
                    \right) = S J_x S,
\end{aligned}
\end{equation}
where $S$ is the diagonal matrix with entries $(1,1,-1)$ and representing a reflection about the plane $(Ox,Oy)$. The Abelian monopole
with unit charge described by $\vec{a}_m$ is now coupled to the new matrix $\tilde{J}_x $. This monopole is thus associated with a Hermitian
matrix which turns out to be a generator, like $J_x$ is, of the rotation subgroup $SO(3)$ of $U(3)$ (see later discussion).
One may note that $J_x= \sqrt{2} \hbar (g_1 + g_6)$  while $\tilde{J}_x= \sqrt{2}\hbar (g_1 - g_6)$ , where the matrices $g_{1,6}$ are
Gell-Mann matrices~\cite{Georgi}. For sake of completeness, we give the 8 Gell-Mann matrices $g_i$ ($1\leq i \leq 8$) in the Appendix.
By changing the laser beams configuration, a $U(3)$ monopole associated with another combination of Gell-Mann matrices could be obtained in principle.

Similarly to the $U(2)$ situation studied in~\cite{Mottonen09}, understanding the topological properties~\cite{Sinha76} of the ground state and excitations
of interacting particles subjected to a $U(3)$ monopole field would certainly lead to new and interesting physics that could be targeted in cold atoms experiments.
More precisely, since the $U(3)$ gauge group is larger than $U(2)$, one expects that different kind of topological charges show up, depending on the symmetry of the ground state.

\section{Spin-orbit coupling}
\label{socsec}
A non-Abelian SO-coupling can alternatively be achieved by considering the laser beams configuration shown in Fig.~\ref{su3sl}
but where all six laser beams are now linearly-polarized plane waves with the following Rabi frequencies:
\begin{equation}
\label{soc}
\begin{aligned}
\Omega_{1,2}^{l}&=\frac{\Omega_{l}}{\sqrt{2}}\sin{\theta_l} e^{\pm i \vec{k}_l\cdot\vec{r}},
\quad \Omega_{3}^{l}=\Omega_{l}\cos{\theta_l} e^{-i\vec{k}_3\cdot\vec{r}},\\
\Omega_{1,2}^{r}&=\frac{\Omega_{r}}{\sqrt{2}}\sin{\theta_r} e^{\pm i \vec{k}_r\cdot\vec{r}}, \quad \Omega_{3}^{r}=\Omega_{r}\cos{\theta_r} e^{i\vec{k}_3\cdot\vec{r}},
\end{aligned}
\end{equation}
($\varphi_r=\varphi_l = \pi/4$). Here we have $\vec{k}_3 \perp \vec{k}_{l,r}$, these vectors being in the plane $(Ox,Oz)$ orthogonal to the quantization axis $Oy$, the polarization states being the same as in Fig.~\ref{su3sl}. Simplifying further to the case $\theta_r=\theta_l= \Theta$, the effective vector potential reads:
\begin{align}
\vec{\mathbf{A}} =& \frac{1}{2} \vec{k}_3 \, J_z + \frac{2\hbar}{\sqrt{2+\tan^2{\Theta}} } (\vec{k}_l \, g_1 + \vec{k}_r \, g_6)\\
=& \frac{1}{2} \vec{k}_3 \, J_z + \frac{2\hbar\vec{k}_l}{\sqrt{2+\tan^2{\Theta}} } (g_1 \pm g_6),
\end{align}
where the last equality has been obtained by also assuming $\vec{k}_r=\pm\vec{k}_l$.
One can generate different types of SO-coupling, in the spirit of what is done in~\cite{Vaishnav2008}
to induce a Rashba or a Linear Dresselhaus SO-coupling. For example, from the expansion of $(\vec{p} - \vec{\mathbf{A}})^2/(2m)$,
we get the SO coupling terms $V_{SO} = \hbar k/(2m) \, v_{SO}$ where
\begin{equation}
\label{soceq}
 \begin{aligned}
  & v_{SO} = (J_z/\hbar) \, p_x + 2\beta(\Theta) (g_1\pm g_6) \, p_z\\
  & v_{SO} =2 \beta(\Theta) (g_1\pm g_6)\, p_x + (J_z/\hbar) \, p_z
 \end{aligned}
\end{equation}
for $(\vec{k}_3, \vec{k}_l) =(k \hat{e}_x, k \hat{e}_z)$ and for $(\vec{k}_3, \vec{k}_l) =(k \hat{e}_z, k \hat{e}_x)$. Here $\beta(\Theta)=2/\sqrt{2+\tan^2{\Theta}}$ and equals unity if $\tan\Theta = \sqrt{2}$, i.~e. if all left (resp. right) Rabi frequencies are equal. One should note that, contrary to the
$U(2)$ scheme, the expansion features the non-Abelian potential $\vec{\mathbf{A}}\cdot \vec{\mathbf{A}}/(2m)$ which adds up to the non-Abelian potential $\boldsymbol{\Phi}$. As one can easily check, this term is not proportional to the identity and therefore should play a role in the dynamics and in the ground state properties of an
interacting system. For $\tan\Theta = \sqrt{2}$, it reads:
\begin{equation}
\frac{\vec{\mathbf{A}}\cdot \vec{\mathbf{A}}}{2m} = \frac{\hbar^2k^2}{4m} \, (\mathbbm{1} - g_4).
\end{equation}

\section{Experimental realization and limitations}
\label{section_Possible_exp_realization}

In Sec.~\ref{Section_U3_monople}, we have shown that non-Abelian $U(3)$-monopole contributions can be obtained using specific laser beams configurations.
More precisely, the common ground state shared by the two tripods, namely ground state level $|3\rangle$, should be coupled to excited states
with Hermite-Gauss laser beams propagating along the same axis. The other ground states are coupled to excited states with Laguerre-Gauss
beams propagating perpendicularly to the Hermite-Gauss beams, see Eqs.~\eqref{u3jx},\eqref{u3njx} and Figs.~\ref{su3sl},\ref{su3nsl}.
Because of these constraints, one cannot solely rely on the laser polarization degrees of freedom and dipole selection rules to independently
and selectively address the different transitions of the 2-tripod scheme. One also needs to apply a strong magnetic field to lift the
Zeeman degeneracy in the ground state and excited state manifolds to get well separated transitions and avoid spurious spontaneous
emission processes from unwanted transitions.

For the $D_1$ line of $^{87}$Rb that we used as an example, this Zeeman degeneracy lifting is even favored by the opposite signs of the Land\'{e} factors
in the ground state ($F_g=2$) and in the excited state ($F_e=1$). Regardless of 
any possible technical issues in generating the required magnetic field, its maximum value is limited by the
hyperfine splitting $\Delta_{hs}\approx 142\Gamma$ of the excited state ($\Gamma = 2\pi \times 5.8$ MHz is the natural line width of the transition),
otherwise the coupling to the other hyperfine manifold $F_e=2$ will start playing a non-negligible role.
In addition, since light-induced gauge potentials originate from photon momentum exchanges with the atoms,
their energy scale is thus of the order of the recoil energy, $E_R = \hbar\omega_R \approx 0.6 \,10^{-3}\hbar\Gamma$.
Therefore, the rate of any residual spontaneous emission from the
off-resonant transitions should be smaller than $\hbar^{-1}E_R$,
which, in the case of $^{87}$Rb, might be difficult to achieve.
Along the same line of thought, it becomes even more challenging to address the effect of the gauge potential on the atoms
dynamics over time scales in the millisecond range and beyond~\cite{Spielman09b}.

\begin{figure}[tb]
\centering
\includegraphics*[width=8cm]{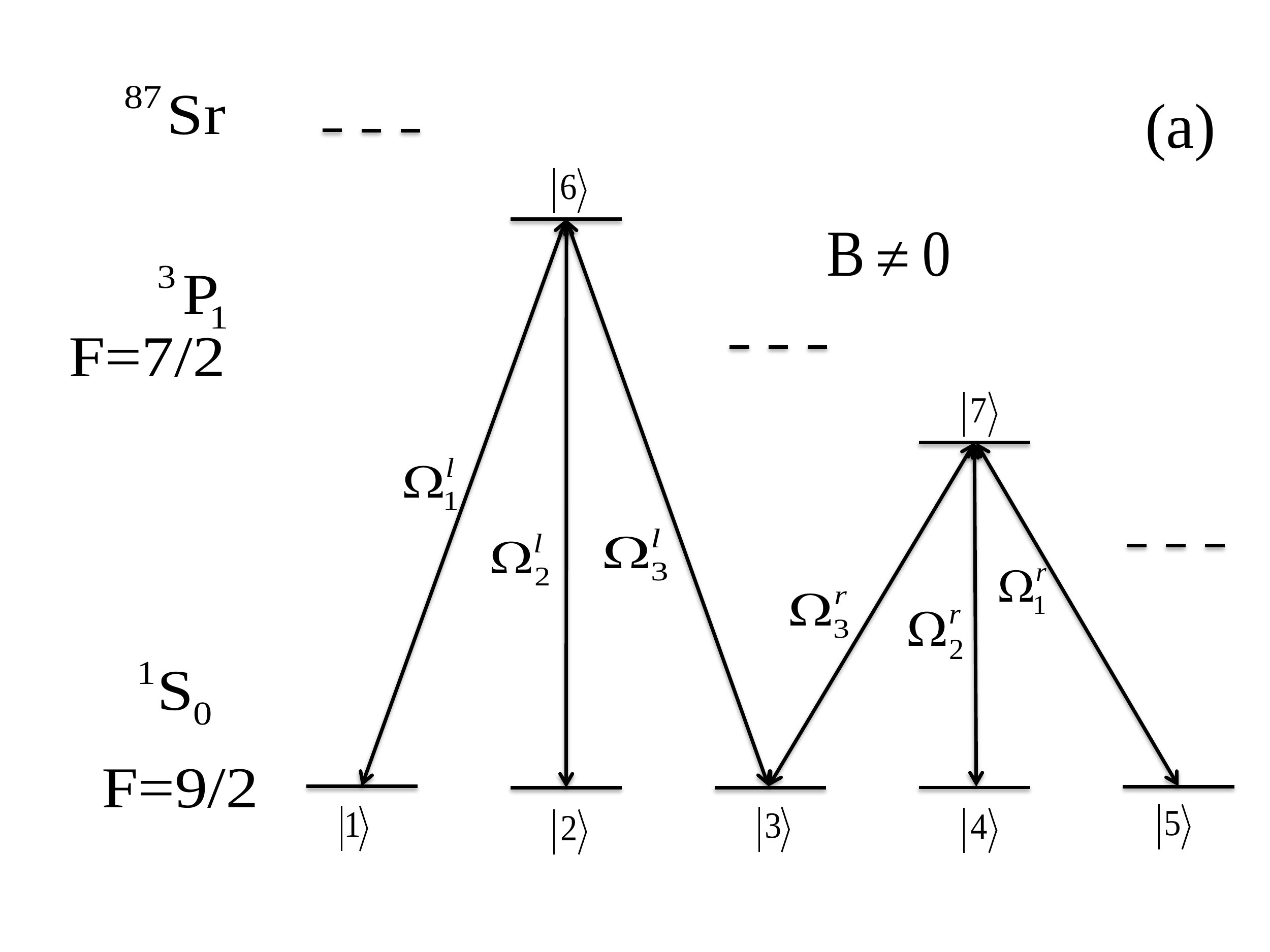}
\includegraphics*[width=8cm]{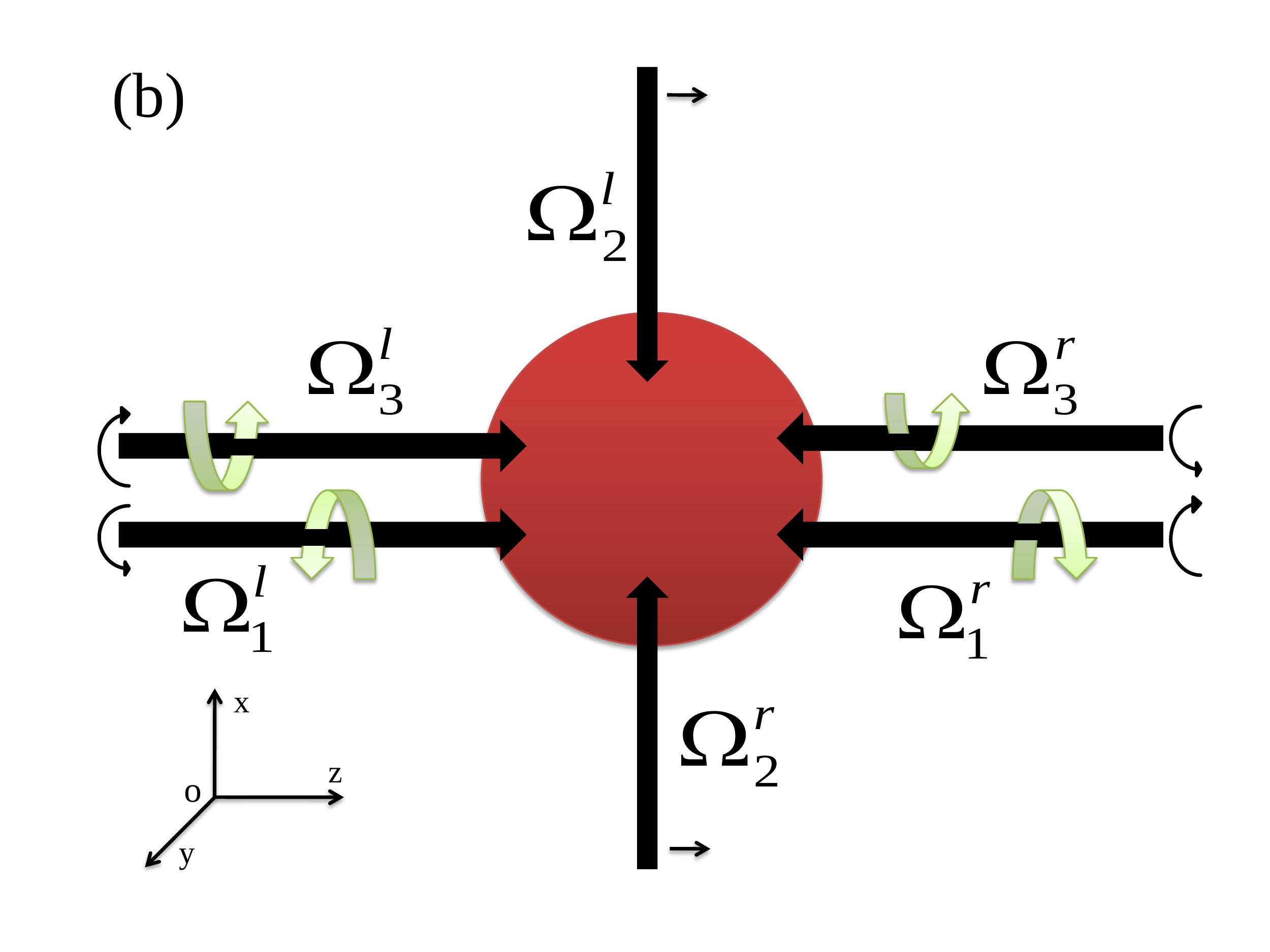}
\caption{\label{dtripod_strontium} [Color online] (a) 2-tripod scheme in the case of $^{87}$Sr atoms. An external magnetic field 
is still applied but, for reasonable magnetic strength, only the Zeeman degeneracy of the excited state manifold is lifted 
(the ground state manifold carries no electronic spin).
Now, contrary to the Rubidium case, the laser fields must have exactly the polarization required by the transition they need to address. T
his imposes the Zeeman field to be applied along the quantization axis $Oz$, the left and right Laguerre-Gauss beams 
to propagate along $Oz$ with opposite circular polarizations (shown by thin black arrows) to address 
the $\sigma_\pm$-transitions. The vertical Hermite-Gauss beams propagate along $Ox$ and are linearly-polarized 
along $Oz$ to address the $\pi$-transitions, see (b).}
\end{figure}

From this point of view, fermionic isotopes of alkaline-earth atoms provide interesting alternatives to Rubidium atoms.
In particular, the hyperfine splitting of the $^3\!P_1$ excited state of the Strontium isotope $^{87}$Sr is $\Delta_{hs} \approx 10^{5}\Gamma$,
at least three orders of magnitude larger than that of alkali atoms.
In this case, the left and right tripod configurations can be driven independently  (see Fig.~\ref{dtripod_strontium}) and
should be well protected from spurious spontaneous emission due to unwanted off-resonant transitions. Furthermore the narrow linewidth ($\Gamma = 7.4$ kHz) of the intercombination line at $689$nm leads to a large Zeeman shift (compared to $\Gamma$) with reasonable magnetic fields of few tens of Gauss. However the $^1\!S_0$ ground state carries no electronic spin, only a nuclear one $I=9/2$. Regarding our laser coupling scheme, the Zeeman shift in the ground state is thus essentially negligible and one has now to solely rely on the polarization states of the beams to address independently and selectively the left and right tripod transitions. More precisely, the Laguerre-Gauss beams addressing the left (resp. right) tripod, and propagating along the $Oz$ axis, must have opposite circular polarizations and should now drive the $1\leftrightarrow 6$ and $3\leftrightarrow 6$ $\sigma_\pm$-transitions (resp. the $3\leftrightarrow 7$ and $5\leftrightarrow 7$ $\sigma_\pm$-transitions). The Hermite-Gauss beams, propagating along the $Ox$ axis, must have a linear polarization, along the $Oz$ axis, and address the $2\leftrightarrow 6$ and $4\leftrightarrow 7$ $\pi$-transitions. One immediately sees that the laser beams configuration proposed in Sec.~\ref{Section_U3_monople}, and generating an $U(3)$-monopole, fails to meet these polarization constraints. Figure~\ref{dtripod_strontium} shows the new laser beams configuration with the appropriate quantization axis and polarization states. In the following Sec.~\ref{section_ae_atoms}, we discuss in some details the properties of the resulting new gauge potentials. In particular, we show that a $U(3)$-monopole with a non-zero charge can still be recovered by using an appropriate gauge transformation.

One may also note that the SO-coupling scheme, depicted in Sec.~\ref{socsec}, can be extended to the case of 
the alkaline-earth atoms, again using appropriate polarization states for the laser beams to address the proper transitions. 
One finds SO-coupling terms similar to those given in~Eq~\eqref{soceq}.

The initial preparation of the atoms in a specific dark state is rather simple. 
Starting from a fully spin-polarized sample, all the lasers connecting the empty Zeeman sub-states of 
the ground level should be initially switched on. Then a slow ramping of the remaining laser fields guarantees 
a full transfer into a given dark state. The final detection can be done by abruptly turning off the laser fields. 
In this case, the dark states are mapped onto the bare spin states. The "flavor" texture of the gas can then be measured 
by using spin-dependent imaging techniques, see Refs.~\cite{Stenger99, Stamper13} for instance. 
Interestingly, we note that the bare states basis has a larger dimension than the dark states basis. It means 
that useful information on the relative phase of the dark states could be extracted as well.

We close this Section by reminding that the gauge field description of the atomic dynamics relies on the assumption of an 
adiabatic evolution which should be fulfilled at any time. This means that the mean Rabi frequency should be much larger than 
any other characteristic frequencies governing the dynamics of the atomic external degrees of freedom. 
In particular one should prevent atoms from going too close to the monopole singularity, where where all laser fields vanish, 
because the adiabatic assumption would break.

\section{alkaline-earth atoms}
\label{section_ae_atoms}

\subsection{$U(3)$ monople}
Taking into account the polarization constraints that alkaline-earth atoms bring into the game, a
laser configuration that corresponds to a realistic experimental situation is the following:

\begin{align}
\Omega_{1,3}^{l}=&\Omega_{0}\frac{\rho}{R} e^{i(kz\mp \varphi)}, \quad \Omega_{2}^{l}=\Omega_{0}\frac{z}{R}e^{ikx}.\\
\nonumber\\
\Omega_{1,3}^{r}=&\Omega_{0}\frac{\rho}{R} e^{-i(kz\mp \varphi)}, \quad \Omega_{2}^{r}=\Omega_{0}\frac{z}{R}e^{-ikx}.
\end{align}

The corresponding non-Abelian vector potential is:
\begin{equation}
\label{su3p}
\begin{aligned}
\vec{A}&=\hbar \, \frac{\cos\theta\sin\theta}{\sqrt{1+2\sin^{2}\theta}}\frac{\hat{e}_{\varphi}}{r}\left(
                      \begin{array}{ccc}
                        0 & 1 & 0 \\
                        1 & 0 & -1\\
                        0 & -1& 0 \\
                      \end{array}
                    \right)\\
&+\hbar \, \frac{(2-\sin^{2}\theta)}{\sin\theta} \frac{\hat{e}_{\varphi}}{r}\left(
                      \begin{array}{ccc}
                        1 & 0 & 0 \\
                        0 & 0 & 0\\
                        0 & 0 & -1 \\
                      \end{array}
                    \right)+ (\cdots),
\end{aligned}
\end{equation}
where $(\cdots)$ represents
terms which do not contribute to the singular part of the radial magnetic field
such as (non-singular) $k$-dependent terms or the ($\varphi$-independent)  $\hat{e}_\theta$ component.
Each term explicitly written in Eq.~\eqref{su3p} correspond to a non-Abelian monopole-like magnetic field,
the first one coupled to $2\hbar(g_1-g_6)$ and the second one to $J_z$ (which is also a linear combination of 
the Gell-Mann matrices of the $SU(3)$ group).
However, each of these terms alone is such that $g(\pi)=g(0)$. They thus each carry a vanishing magnetic
charge according to Eq.~\eqref{eq_charge} and the present configuration seems not, strictly speaking, to produce
a true magnetic monopole. However, let us apply the gauge transformation $U=e^{i\theta \tilde{J}_y/\hbar}$ where
\begin{equation}
\tilde{J}_y= \frac{\hbar}{\sqrt{2}} \, \left(
\begin{array}{ccc}
0 & -i & 0 \\
 i & 0 & i\\
0 & -i& 0 \\
\end{array}\right) = S J_y S.
\end{equation}
Since $\tilde{J}_a = S J_a S$ ($a=x,y,z$) satisfies the usual angular momentum commutation relations 
$[\tilde{J}_a, \tilde{J}_b] = \mathrm{i}\hbar \varepsilon_{abc} \tilde{J}_c$ (note that $\tilde{J}_z = S J_z S = J_z$), 
one can use the rotation algebra and easily compute the transformed monopole gauge field 
$\vec{\mathbf{A}}{}'=U\vec{\mathbf{A}}U^\dagger+\mathrm{i}\hbar \, U \vec{\nabla} U^\dagger$:
\begin{equation}
\begin{aligned}
\vec{\mathbf{A}}{}^{'}&=(2-\sin^{2}\theta) (\tilde{J}_x + \cot\theta \, J_{z})\frac{\hat{e}_{\varphi}}{r}\\
&+\frac{\cos\theta\sin^2\theta}{\sqrt{1+2\sin^{2}\theta}} \left(\cot\theta \, \tilde{J}_x + \tilde{J}_y -J_z \right)\frac{\hat{e}_{\varphi}}{r}\\
&+ \ (...),
\end{aligned}
\end{equation}
where again $(\cdots)$ represents the terms which do not contribute to the singular part of the radial magnetic field.
The only term describing a true $U(3)$-monopole is
\begin{align}
\vec{\mathbf{A}}{}^{'}_m &= \frac{2 \cos\theta}{\sin\theta} \, \frac{\hat{e}_{\varphi}}{r} \, J_z \\\vec{\mathbf{B}}{}^{'}_m &= -2 \,\frac{\hat{e}_r}{r^2} \, J_z
\end{align}
with non-vanishing charge $Q=-2$.

\subsection{Gauge transformations and magnetic charge}

The previous situation is similar to the relationship between the $U(2)$ $\textrm{'t Hooft}$-Polyakov monopole and 
the $U(1)$ Dirac monopole~\cite{Ryder}. A non-Abelian gauge transformation can be used to fully remove the string 
singularity of the Dirac monopole along the negative $z$ axis by transferring it to the associated Higgs field. 
In the process, it is the total charge of the monopole and of the Higgs field that is conserved.  For example, 
in Ref.~\cite{Mottonen09}, the Authors compute the ground state properties of a Bose-Einstein condensate 
subjected to a $U(2)$-monopole (namely an Abelian one coupled to $\sigma_x$). They apply a gauge transformation to remove the string singularity
(along the full $z$ axis) to get a new non-Abelian gauge field but with a vanishing magnetic charge. This is the 
kind of situation we face here where we find a $U(3)$-monopole with a vanishing charge in one gauge and with a non-vanishing charge in another gauge.

Let us illustrate a bit further the modification of the effective magnetic charge of the non-Abelian monopole when 
a unitary change of the DS basis set is performed by discussing the usual tripod scheme studied in~\cite{Ruseckas05} and~\cite{Dalibard11}. 
It is obtained from our 2-tripod scheme by setting $\Omega^r_i =0$. One can build a DS on
the Zeeman states $|1\rangle$ and $|2\rangle$ coupled to the excited state $\vert 6\rangle$ by the $\sigma_+$ 
and $\pi$ Laguerre-Gauss beams. The DS basis set is then:
\begin{equation}
 \label{DS2}
 \begin{aligned}
  |D_1\rangle =& \frac{1}{\sqrt{2}}\left[e^{iS_{31}}|1\rangle - e^{iS_{32}}|2 \rangle\right]\\
  |D_2\rangle =&\frac{1}{\sqrt{2}} \cos\theta \left[ e^{iS_{31}}\vert 1 \rangle +
                       e^{iS_{32}}\vert 2\rangle \right]   - \sin{\theta} \, \vert 3 \rangle.
 \end{aligned}
\end{equation}
Dropping the $k$-dependent part for simplicity, these DS generate the usual $U(2)$-monopole gauge
field, $\cos{\theta} \, \sigma_x  \, \hat{e}_{\phi}/(r\sin\theta)$ which has an effective magnetic charge $Q=-1$.

If, on the contrary,
one chooses to build a DS on the Zeeman states $|1\rangle$ and $|3\rangle$ coupled to the excited
state by the $\sigma_+$ Laguerre-Gauss beam and by the $\sigma_-$ Hermite-Gauss beam, one instead gets:
\begin{equation}
 \label{DS2p}
 \begin{aligned}
  |D'_1\rangle =& \frac{1}{\sqrt{1+\cos^2\theta}}\left[\sqrt{2}\cos\theta e^{iS_{21}}|1\rangle - \sin\theta e^{iS_{23}}|3 \rangle\right]\\
  |D'_2\rangle =&\frac{1}{2\sqrt{1+\cos^2\theta}}\left[\sqrt{2} \sin^2\theta e^{iS_{21}}\vert 1 \rangle +\right.\\
  &\left.
                      2\sin\theta\cos\theta e^{iS_{23}}\vert 3\rangle - \sqrt{2}(1+\cos^2\theta)\vert 2 \rangle\right].
 \end{aligned}
\end{equation}
Dropping again the terms which
do not contribute to the singular part of the radial magnetic field, the new vector potential is:
\begin{equation}
\begin{aligned}
 \vec{\mathbf{A}}{}' =
       \left[\openone+\frac{\cos\theta\sin^2\theta}{1+\cos^2\theta}\sigma_x +\frac{2\cos^2\theta}{1+\cos^2\theta}\sigma_z
	\right]\frac{\hbar \, \hat{e}_{\phi}}{r\sin\theta},
      \end{aligned}
\end{equation}
where $\sigma_x$ and $\sigma_z$ are the standard Pauli matrices. These expressions bear some resemblance 
to the $U(3)$ situation, see Eq.\eqref{su3p}. Here too, each term gives rise to a vanishing magnetic charge $Q=0$. On the other
hand, the transformation from the $|D'_a\rangle$ basis to the $|D_a\rangle$ basis ($a=1,2$) is given by the unitary matrix
\begin{equation}
 U=\frac{e^{\mathrm{i}S_{23}}}{\sqrt{1+\cos^2\theta}}\left(\begin{array}{cc}
        \cos\theta & 1\\
        1 & -\cos\theta
       \end{array}\right).
\end{equation}
This shows that the gauge fields $\vec{\mathbf{A}}{}'$ and $\vec{\mathbf{A}}$ are simply related by the gauge
transformation $\vec{\mathbf{A}}{}'=U\vec{\mathbf{A}}U^{\dagger}+\mathrm{i}\hbar U\vec{\nabla} U^{\dagger}$.
Therefore, the two gauge fields correspond to the same physical situation, a $U(2)$-monopole, but
with $\vec{\mathbf{B}}{}' = U \vec{\mathbf{B}} U^{\dagger}$ despite the fact that the effective magnetic charges are different.

This is somehow what we have obtained for the $U(3)$ case above, i.e. the physics of a $U(3)$-monopole but
from an unusual gauge perspective. However it is important to note that, contrary to the $U(2)$ case, the $U(3)$
gauge potentials obtained with the laser configuration used for the alkaline-earth atoms and the gauge potentials
obtained with the laser configuration used for the Rubidium case are \emph{not} related by a gauge transformation.
Indeed the DS obtained in one scheme are linearly independent from the DS obtained with the other scheme since they
span different subspaces of the full Hilbert space. Therefore they cannot be related by a (position-dependent) $3\times3$
unitary matrix, as required for a proper gauge transformation, only a larger one in the full Hilbert space can.

\section{conclusion}

In this paper we have proposed a workable experimental scheme, the so-called 2-tripod configuration,
to generate non-Abelian $U(3)$ artificial static gauge fields for cold atomic gases. Our scheme only relies
on one-photon resonant transitions and gives rise to three degenerate dark states. We have given the laser beams
configurations for both alkali and alkaline-earth atoms and explained how to generate a $U(3)$-monopole or a $U(3)$
spin-orbit coupling. Future work includes the study of the properties of the ground state and excitations of spinor
condensates subjected to such $U(3)$ gauge potentials, in particular from a topological point of view (spin textures, etc).

\medskip
The Centre for Quantum Technologies is a Research Centre of Excellence funded by
the Ministry of Education and National Research Foundation of Singapore.

\appendix
\section{General expression for the scalar potential}
The entries for the scalar potential Hermitian matrix $\boldsymbol{\Phi}$ read:

\label{scalpot}
\begin{equation}
\begin{aligned}
\Phi_{11}=&\frac{1+\cot^{2}\theta_{r}}{1+\cot^{2}\theta_{l}+\cot^{2}\theta_{r}}
\biggl[\frac{1}{4}\sin^{2}(2\phi_{l})(\vec{\nabla} S^{l}_{12})^{2}+(\vec{\nabla} \phi_{l})^{2}\biggr]\\
\Phi_{33}=&\frac{1+\cot^{2}\theta_{l}}{1+\cot^{2}\theta_{l}+\cot^{2}\theta_{r}}
\biggl[\frac{1}{4}\sin^{2}(2\phi_{r})(\vec{\nabla} S^{r}_{12})^{2}+(\vec{\nabla} \phi_{r})^{2}\biggr]\\\nonumber
\end{aligned}
\end{equation}

\begin{equation}
\begin{aligned}
\Phi_{22}=& \frac{ \cot^{2}\theta_{l} (1+\cot^{2}\theta_{r}) } {(1+\cot^{2}\theta_{l}+\cot^{2}\theta_{r})^{2}}
\left(\cos^{2}\phi_{l}\vec{\nabla} S^{l}_{13}+\sin^{2}\phi_{l}\vec{\nabla} S^{l}_{23}\right)^{2}\\
+&\frac{\cot^{2}\theta_{r} (1+\cot^{2}\theta_{l})}  {(1+\cot^{2}\theta_{r}+\cot^{2}\theta_{l})^{2}}
\left(\cos^{2}\phi_{r}\vec{\nabla} S^{r}_{13}+\sin^{2}\phi_{r}\vec{\nabla} S^{r}_{23}\right)^{2}\\
-&2\frac{\cot^{2}\theta_{l}\cot^{2}\theta_{r}}{(1+\cot^{2}\theta_{l}+\cot^{2}\theta_{r})^{2}}
\left(\cos^{2}\phi_{r}\vec{\nabla} S^{r}_{13}+\sin^{2}\phi_{r}\vec{\nabla} S^{r}_{23}\right)\\
&\phantom{2\frac{\cot^{2}\theta_{l}\cot^{2}\theta_{r}}{(1+\cot^{2}\theta_{l}+\cot^{2}\theta_{r})^{2}}}\times
\left(\cos^{2}\phi_{l}\vec{\nabla} S^{l}_{13}+\sin^{2}\phi_{l}\vec{\nabla} S^{l}_{23}\right)\\
+&\left(\vec{\nabla} \frac{\cot\theta_{l}}{(1+\cot^{2}\theta_{l}+\cot^{2}\theta_{r})^{1/2}}\right)^{2}\\
+&\left(\vec{\nabla} \frac{\cot\theta_{r}}{(1+\cot^{2}\theta_{l}+\cot^{2}\theta_{r})^{1/2}}\right)^{2}\\
+&\left(\vec{\nabla} \frac{1}{(1+\cot^{2}\theta_{l}+\cot^{2}\theta_{r})^{1/2}}\right)^{2}\nonumber
\end{aligned}
\end{equation}

\begin{equation}
\begin{aligned}
\Phi_{13}=&-\frac{\cot\theta_{l}\cot\theta_{r}}{1+\cot^{2}\theta_{l}+\cot^{2}\theta_{r}}
\left(\frac{1}{2}\sin(2\phi_{l})\vec{\nabla} S^{l}_{12}-i\vec{\nabla} \phi_{l}\right)\times\\
&\phantom{-\frac{\cot\theta_{l}\cot\theta_{r}}{1+\cot^{2}\theta_{l}+\cot^{2}\theta_{r}}}
\left(\frac{1}{2}\sin(2\phi_{r})\vec{\nabla} S^{r}_{12}+i\vec{\nabla} \phi_{r}\right)\nonumber
\end{aligned}
\end{equation}

\begin{equation}
\begin{aligned}
\Phi_{12}=& i\vec{\nabla}\left(\frac{\cot\theta_{l}}{(1+\cot^{2}\theta_{l}+\cot^{2}\theta_{r})^{1/2}}\right)
(\frac{1}{2}\sin(2\phi_{l})\vec{\nabla} S^{l}_{12} - i\vec{\nabla} \phi_{l})\nonumber\\
&+\frac{\cot\theta_{l} (1+\cot^{2}\theta_{l})}{(1+\cot^{2}\theta_{l}+\cot^{2}\theta_{r})^{3/2}}
(\frac{1}{2}\sin(2\phi_{l})\vec{\nabla} S^{l}_{12}-i\vec{\nabla} \phi_{l})\\
&\times(\cos^{2}\phi_{l}\vec{\nabla} S^{l}_{13}+\sin^{2}\phi_{l}\vec{\nabla} S^{l}_{23})\nonumber\\
&-\frac{\cot\theta_{l}\cot^{2}\theta_{r}}{(1+\cot^{2}\theta_{l}+\cot^{2}\theta_{r})^{3/2}}
(\frac{1}{2}\sin(2\phi_{l})\vec{\nabla} S^{l}_{12}-i\vec{\nabla} \phi_{l}) \\
&\times(\cos^{2}\phi_{r}\vec{\nabla} S^{r}_{13}+\sin^{2}\phi_{r}\vec{\nabla} S^{r}_{23})
\end{aligned}
\end{equation}

\begin{equation}
\begin{aligned}
\Phi_{32}=&i\vec{\nabla}\left(\frac{\cot\theta_{r}}{(1+\cot^{2}\theta_{l}+\cot^{2}\theta_{r})^{1/2}}\right)
(\frac{1}{2}\sin(2\phi_{r})\vec{\nabla} S^{r}_{12} - i\vec{\nabla} \phi_{r})\nonumber\\
&+\frac{\cot\theta_{r} (1+\cot^{2}\theta_{l})}{(1+\cot^{2}\theta_{l}+\cot^{2}\theta_{r})^{3/2}}
(\frac{1}{2}\sin(2\phi_{r})\vec{\nabla} S^{r}_{12}-i\vec{\nabla} \phi_{r})\\
&\times(\cos^{2}\phi_{r}\vec{\nabla} S^{r}_{13}+\sin^{2}\phi_{r}\vec{\nabla} S^{r}_{23})\nonumber\\
&-\frac{\cot\theta_{r}\cot^{2}\theta_{l}}{(1+\cot^{2}\theta_{r}+\cot^{2}\theta_{r})^{3/2}}
(\frac{1}{2}\sin(2\phi_{r})\vec{\nabla} S^{r}_{12}-i\vec{\nabla} \phi_{r})\\
&\times(\cos^{2}\phi_{l}\vec{\nabla} S^{l}_{13}+\sin^{2}\phi_{l}\vec{\nabla} S^{l}_{23})
\end{aligned}
\end{equation}

\section{The Gell-Mann matrices}
The set of Gell-Mann matrices $g_i$ ($1 \leq i \leq 8$) is one possible representation of the infinitesimal 
generators of the special unitary group SU(3). An important representation features the $3\times 3$ Hermitian $\lambda$-matrices $g_i =\lambda_i/2$ with:
\begin{align}
\lambda_1 &= \left(\begin{array}{ccc}
                        0 & 1 & 0 \\
                        1 & 0 & 0\\
                        0 & 0& 0 \\
                      \end{array}
                    \right)
                    \quad \lambda_2 = \left(\begin{array}{ccc}
                        0 & - \mathrm{i} & 0 \\
                        \mathrm{i}  & 0 & 0\\
                        0 & 0& 0 \\
                      \end{array}\right) \nonumber \\
\lambda_3 &= \left(\begin{array}{ccc}
                        1 & 0& 0 \\
                        0 & -1 & 0\\
                        0 & 0& 0 \\
                      \end{array}\right)
                      \quad
                      \lambda_4 = \left(\begin{array}{ccc}
                        0 & 0 & 1 \\
                        0 & 0 & 0\\
                        1 & 0& 0 \\
                      \end{array}
                    \right)
                    \nonumber \\
\lambda_5 &= \left(\begin{array}{ccc}
                        0 & 0 & - \mathrm{i} \\
                        0& 0 & 0\\
                        \mathrm{i} & 0& 0 \\
                      \end{array}\right)
                      \quad \lambda_6 = \left(\begin{array}{ccc}
                        0 & 0 & 0 \\
                        0 & 0 & 1\\
                        0 & 1& 0 \\
                      \end{array}
                    \right) \nonumber \\
\lambda_7 &= \left(\begin{array}{ccc}
                        0 & 0 & 0 \\
                        0 & 0 & -\mathrm{i} \\
                        0 & \mathrm{i} & 0 \\
                      \end{array}
                    \right)
                    \quad \lambda_8 = \frac{1}{\sqrt{3}}\, \left(\begin{array}{ccc}
                        1 & 0 & 0\\
                        0 & 1 & 0\\
                        0 & 0& -2 \\
                      \end{array}\right).   \nonumber
\end{align}

\section{Scalar potential for the $U(3)$ case}
The scalar potential for the laser scheme described by Eq.~\eqref{u3jx} ($+$ sign) or by Eq.~\eqref{u3njx} ($-$ sign) reads:
\begin{equation}
\begin{aligned}
\boldsymbol{\Phi}_\pm=\frac{\hbar^{2}}{2m} \left[\right. &\left.(\frac{1}{2r^{2}\sin^{2}\theta}+\frac{1}{2r^{2}})\left(
                                                        \begin{array}{ccc}
                                                          1 & 0 & 0 \\
                                                          0 & 0 & 0 \\
                                                          0 & 0 & 1 \\
                                                        \end{array}
                                                      \right)\right.\\
                               +&(\frac{1}{r^2}+\frac{k^{2}}{2}\sin^{2}2\theta)\left(
                                                      \begin{array}{ccc}
                                                       0 & 0 & 0 \\
                                                       0 & 1 & 0 \\
                                                       0 & 0 & 0 \\
                                                      \end{array}
                                                      \right)\\
& \mp\frac{\cos^{2}\theta}{2r^2\sin^{2}\theta}\left(
                                              \begin{array}{ccc}
                                                0 & 0 & 1 \\
                                                0 & 0 & 0 \\
                                                1 & 0 & 0 \\
                                              \end{array}
                                            \right)\\
&\left.-\frac{\sqrt{2}k}{4r}\sin 2\theta\sin\varphi\left(
                                                                                           \begin{array}{ccc}
                                                                                             0 & 1 & 0 \\
                                                                                             1 & 0 & \pm1 \\
                                                                                             0 & \pm1 & 0 \\
                                                                                           \end{array}
                                                                                         \right)\right].
\end{aligned}
\end{equation}


\begin{thebibliography}{11}
\bibitem{stringcharge} One may note that the magnetic charge associated to the string is simply $Q_s=-Q_m$.
\end{thebibliography}
\end{document}